\documentclass[apjl]{emulateapj}

\usepackage{hyperref}
\usepackage{amsmath}
\usepackage{amssymb}
\usepackage{graphicx}
\usepackage{color}
\usepackage{natbib}

\def\apj{Astrophysical Journal}                
\def\apjl{Astrophysical Journal}             

\def\mnras{MNRAS}            
\def\prd{Phys.~Rev.~D}       
\def\nat{Nature}              
\def\aap{A\&A}                

\def\be{\begin{equation}}
\def\ee{\end{equation}}
\def\ba{\begin{eqnarray}}
\def\ea{\end{eqnarray}}

\begin{document}

\title{The orbit of GW170817 was inclined by less than 28 degrees to the line of sight}
\shorttitle{Inclination of GW170817}

\author{Ilya Mandel\altaffilmark{1}}
\affil{$^1$Institute of Gravitational Wave Astronomy and School of Physics and Astronomy, University of Birmingham, Birmingham, B15 2TT, United Kingdom}
\email{imandel@star.sr.bham.ac.uk}

\begin{abstract}
We combine the gravitational-wave measurement of the effective distance to the binary neutron star merger GW170817, the redshift of its host galaxy NGC 4993, and the latest Hubble constant measurement from the Dark Energy Survey to constrain the inclination between the orbital angular momentum of the binary and the line of sight to $18 \pm 8$ degrees (less than 28 degrees at 90\% confidence).  This provides a complementary constraint on models of potential afterglow observations.
\end{abstract}


\maketitle

\section{Introduction}

Gravitational waves from the coalescence of two neutron stars, GW170817, were detected by the advanced LIGO \citep{AdvLIGO} and Virgo \citep{AdvVirgo} gravitational-wave observatories on 17 August, 2017 \citep{GW170817}.  A short gamma ray burst was observed by Fermi less than two seconds later \citep{GW170817:GRB}.  These detections initiated a campaign of electromagnetic observations which identified the transient source and localized it to the host galaxy NGC 4993 \citep[][and references therein]{GW170817:MMA}.

For a nearly face-on binary located nearly overhead a gravitational-wave detector, the gravitational-wave signal amplitude  scales as $\cos{\iota} / D_L$, where $D_L$ is the luminosity distance to the source and $\iota$ is the inclination angle to the line of sight (0 for a face-on and 180 degrees for a face-off binary).   Gravitational-wave observations measure the signal amplitude to a fractional accuracy of $\sim 1/\rho$, where the signal-to-noise ratio $\rho$ is approximately  32 for GW170817 \citep{GW170817}.  However, the inclination-distance degeneracy prevents a precise inclination measurement, and the viewing angle can only be constrained to be below $55$ degrees from gravitational-wave observations alone \citep{GW170817}.

This inclination-distance degeneracy, inherent in gravitational-wave inference \citep[e.g.,][]{Veitch:2012}, can be broken with the aid of an independent distance measurement.  The observed recession velocity of the host galaxy, NGC 4993 \citep[e.g.,][]{Hjorth:2017}, combined with a precise value of the Hubble constant, provides such a measurement.

The most recent measurement of the Hubble constant by the Dark Energy Survey (DES) team yields a value of $H_0=67.2^{+1.2}_{-1.0}$ km/s/Mpc \citep{DES:2017}.  This is very similar to the $H_0$ value inferred from Planck data \citep{Planck:2015}.  The DES team also combines results with four other statistically independent $H_0$ measurements -- Planck and SPTpol \citep{Henning:2017} which use independent CMB information, SH0ES \citep{Riess:2016} which is based on standard candles provided by type IA supernovae, and H0LiCOW \citep{Bonvin:2017} which uses time delays between images of strongly lensed variable quasars -- to yield the very precise value of $H_0 = 69.1^{+0.4}_{-0.6}$ km/s/Mpc \citep{DES:2017}. 

\citet{GW170817:H0} combined the LIGO/Virgo distance data with the observed redshift of the host galaxy NGC 4993 to obtain an independent measurement of $H_0$.  Here, we reverse their approach: we take the data behind Figure 2 of \citet{GW170817:H0}, which includes posterior samples in the two-dimensional $H_0$--$\cos{\iota}$ space, and resample them according to the tighter, independent $H_0$ measurements in order to obtain a posterior distribution on $\iota$, where we map $\iota$ onto the range $0 \le \iota \le 90$ degrees.  

\section{Results}

\begin{figure}
\vspace{-1in}
    		\includegraphics[trim={0cm 0cm 0cm 0cm},clip,scale=0.4]{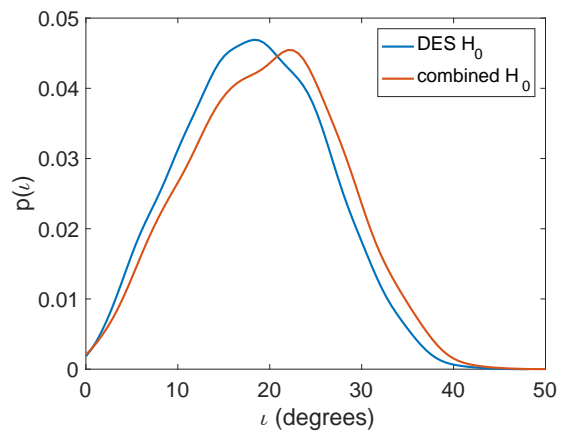}
\vspace{-1in}		
    		\caption{Probability distribution on the orbital inclination angle $\iota$ of GW170817, obtained by combining the distance and inclination inferred from the gravitational-wave signal, the host galaxy redshift, and an independent measurement of the Hubble constant.  The blue curve is based on the DES measurement of $H_0$, while the red curve is based on a combination of 5 statistically independent $H_0$ measurements \citep{DES:2017}.}
\end{figure}

In the Bayesian framework, we marginalize the joint posterior probability distribution on $\iota$ and $H_0$ given both the gravitational-wave data $\vec{d}_\mathrm{GW}$ and independent data $\vec{d}_{H_0}$ that provides improved knowledge $H_0$:
\begin{equation}
p(\iota | \mathrm{data}) = \int p(\iota, H_0 | \vec{d}_\mathrm{GW}, \vec{d}_{H_0}) dH_0.
\end{equation}
We use the existing joint posterior computed from gravitational-wave observations by \citet{GW170817:H0}
\begin{equation}
p(\iota, H_0 | \vec{d}_\mathrm{GW}) \propto \pi_\mathrm{LVC}(\iota) \pi_\mathrm{LVC}(H_0) p(\vec{d}_\mathrm{GW}|\iota, H_0),
\end{equation}
where $\pi_\mathrm{LVC}$ denote the priors used by \citet{GW170817:H0} with $\pi_\mathrm{LVC} (H_0) \propto 1/H_0$.  Therefore,
\begin{eqnarray}
p(\iota, H_0 | \vec{d}_\mathrm{GW}, \vec{d}_{H_0}) & \propto & \pi_\mathrm{LVC}(\iota) p(H_0|\vec{d}_{H_0}) p(\vec{d}_\mathrm{GW}|H_0,\iota) \nonumber\\
& \propto & p(\iota, H_0 | \vec{d}_\mathrm{GW}) \frac{p(H_0|\vec{d}_{H_0})}{\pi_\mathrm{LVC}(H_0)},
\end{eqnarray}
where $p(H_0|\vec{d}_{H_0})$ is the distribution of $H_0$ inferred from independent observations.

Figure 1 shows the posterior on $\iota$ re-weighted by the $H_0$ measurements from \citet{DES:2017}.  Specifically, we re-weighted the data from \citet{GW170817:H0} by the ratio $p(H_0|\vec{d}_{H_0}) / \pi_\mathrm{LVC} (H_0)$.  We approximated the DES measurement $p(H_0|\vec{d}_{H_0})$ as a normal distribution with mean 67.3 km/s/Mpc and standard deviation 1.1 km/s/Mpc.  The inclination angle $\iota$ is constrained to be $18 \pm 8$ degrees with a 90\%-confidence upper limit of 28 degrees.

If instead the combined $H_0$ value from five measurements is used \citep{DES:2017}, approximated as a normal distribution with mean 69.0 km/s/Mpc and standard deviation 0.5 km/s/Mpc, we obtain an inclination angle $\iota = 20 \pm 8$ degrees with a 90\%-confidence upper limit of 30 degrees.  These constraints are summarized in table 1.

Our constraint on low inclination angles is not informative: as $\iota \to 0$, the inferred posterior probability distribution on $\iota$ is consistent with the prior, $p(\iota) \propto \sin{\iota}$, which disfavors very small inclination angles.  \citet{Pian:2017} point out that the a priori probability for $\iota < 26$ degrees is only 10\% \footnote{The a priori probability for $\iota < 26^\circ$ rises to 30\% if selection effects from gravitational-wave searches are included, due to mild on-axis beaming of gravitational waves.}.  However, as pointed out by \citet{Guidorzi:2017}, low values of $\iota$ are ruled out by electromagnetic observations, which indicate that the Earth is neither in the jet nor too close to the jet, given the $\gtrsim 10$-day delay times before X-ray and radio afterglows were detected \citep{Troja:2017,Hallinan:2017}.  

\begin{table}
\begin{tabular}{lcc}
$H_0$ source  & $H_0$ [km s$^{-1}$ Mpc$^{-1}$]  & $\iota$ [degrees] \\
\hline
DES-only\footnote{\citep{DES:2017}} & $67.2^{+1.2}_{-1.0}$ & $18 \pm 8$ (28)\\
DES+$v_\mathrm{pec}$\footnote{$H_0$ as above, with 100 km/s peculiar velocity offset assumed} & & (31)\\
Combined$^\mathrm{a}$ & $69.1^{+0.4}_{-0.6}$& $20 \pm 8$ (30)\\
SH0ES\footnote{\citep{Riess:2016}} & $73.24 \pm 1.74$ & $25 \pm 8$ (35)\\
\hline
\end{tabular}
\caption{The inferred value of the inclination angle $\iota$ (with 90\% upper limits in parentheses) for each of the $H_0$ observations as described in the text.}
\end{table}

\section{Discussion}

The \citet{DES:2017} and \citet{Planck:2015} values of the Hubble constant are significantly lower than the value inferred from type Ia supernova and the local distance ladder, $H_0 = 73.24 \pm 1.74$ km/s/Mpc \citep{Riess:2016}.  The latter value would yield $\iota = 25 \pm 8$ degrees with a 90\%-confidence upper limit of 35 degrees.  However, while the Planck value is based on cosmic microwave background measurements and its discrepancy with the \citet{Riess:2016} value could conceivably be due to a failing of the standard $\Lambda$ CDM cosmology, the DES value is ultimately based on low-redshift galaxy clustering and weak lensing observations, but without the potential systematics inherent in calibrating a distance ladder.

The peculiar velocity of NGC 4993 relative to the Hubble flow can affect the conversion of its redshift into distance.  \citet{GW170817:H0} consider a variation in which the uncertainty on the peculiar velocity is increased to 250 km/s from the nominal 150 km/s; this has virtually no effect (see their extended table I).  We also consider a systematic shift in the peculiar velocity by 100 km/s in addition to this velocity uncertainty.  This corresponds to a 3\% shift in the velocity, and hence a 3\% change in the distance for a given value of $H_0$.  The resulting 3\% shift in $\cos{\iota}$ could move the 90\% upper limit on $\iota$ from 28 to 31 degrees. 

Our maximum inclination angle constraint is significantly tighter than the constraint from gravitational-wave measurements alone, less than 55 degrees at 90\% confidence \citep{GW170817}\footnote{A similar constraint to the one obtained here was made by assuming the Planck value of $H_0$ by \citet{GW170817}.}.  This complementary constraint can aid in the interpretation of the electromagnetic transient associated with this binary neutron star merger.  In particular, it strongly limits the available parameter space for models of observed electromagnetic signatures, ruling out several proposals.  For example, we can rule out the preferred model of \citet{Kim:2017}, which explains radio observations by appealing to an observing angle of 41 degrees.   The preferred `top-hat' jet model of \citet{Lazzati:2017} with the same observing angle can also be ruled out, though their structured jet model is consistent with the inclination constraint presented here.  The model of \citet{Evans:2017}, which prefers an observing angle of $\approx 30$ degrees, is only marginally consistent with the maximum allowed inclination value.  The models of \citet{Nicholl:2017,Troja:2017,Alexander:2017,Margutti:2017,Perego:2017,Mooley:2017} are consistent with the constraint presented here for only part of their parameter space; adding this additional constraint could allow for more precise estimates of other free parameters in these models.  Other models consistent with the maximum inclination value presented here include \citet{Haggard:2017,Fraija:2017}.

As another example application, we can translate the threshold on $\iota$ into a constraint on the jet energy and the density of the surrounding material.  The afterglow peak is expected at
\begin{equation}\label{tpeak}
t_\mathrm{peak} \sim 70 \left(\frac{E}{10^{51}\,\mathrm{erg}}\right)^{1/3} \left(\frac{n}{1\,\mathrm{cm}^{-3}}\right)^{-1/3} \theta_\mathrm{obs}^2\; \; \mathrm{days}
\end{equation}
after the merger \citep{Granot:2017afterglow}, where $E$ is the jet kinetic energy, $n$ is the interstellar medium density, and $\theta_\mathrm{obs}$ is the observing angle in radians.  We assume that the jet is perpendicular to the binary's orbital plane, so $\theta_\mathrm{obs}=\iota$.  The constraint on $\iota$ provided here then yields
\begin{equation}
\frac{E}{10^{50}\,\mathrm{erg}} \frac{10^{-4}\,\mathrm{cm}^{-3}}{n} \gtrsim \left(\frac{t_\mathrm{peak}}{170\,\mathrm{days}}\right)^3 \left(\frac{\iota}{28^\circ}\right)^{-6}.
\end{equation}
The isotropic-equivalent energy of the jet, $E_\textrm{iso} \sim 2 E / \theta_0^2$ where $\theta_0$ is the jet opening angle, has been observed to range from $3 \times 10^{49}$ to $3 \times 10^{52}$ erg, while the inferred circum-merger density $n$ ranges from $\sim 10^{-4}$ to $\sim 1$ cm$^{-3}$ in on-axis events \citep{Fong:2015}.  The jet opening angle has been estimated from jet breaks and from the requirement of matching short gamma ray bursts rates to binary neutron star merger rates to be $\theta_0 \sim 10^\circ$ \citep{FongBerger:2013,Fong:2015}.  Therefore, the observations of a continuing afterglow lasting beyond 100 days after the merger \citep{Mooley:2017,Ruan:2017,Levan:2017GCN,Lazzati:2017} indicate that the merger happened in a very low-density environment, $n \lesssim 10^{-4}$ cm$^{-3}$.  Alternatively, they could indicate that there is limited or no sideways expansion of the jet, which would increase the peak time for a given choice of $E$ and $n$ by a factor of $(\theta_\mathrm{obs}/\theta_0)^{2/3}$ relative to Eq.~(\ref{tpeak}), therefore relaxing the constraint on the maximum $n$ obtained here by a factor of $(\theta_\mathrm{obs}/\theta_0)^2$; even then, the rise of the afterglow should not extend beyond $\sim 550$ days after merger for $n \gtrsim 10^{-4}$ cm$^{-3}$.

\acknowledgements
We thank Christopher Berry, Jens Hjorth and Nial Tanvir for useful discussions.  We particularly thank Will Farr for contributing to the formulation of this work; unfortunately, he did not wish to be an author because of restrictions required by  the LIGO Scientific Collaboration policies. IM is partially supported by STFC.

\bibliographystyle{hapj}

\begin{thebibliography}{32}
\expandafter\ifx\csname natexlab\endcsname\relax\def\natexlab#1{#1}\fi

\bibitem[{{Aasi} {et~al.}(2015){Aasi}, {Abbott}, {Abbott}, {Abbott},
  {Abernathy}, {Ackley}, {Adams}, {Adams}, {Addesso}, \& et~al.}]{AdvLIGO}
{Aasi}, J. {et~al.} 2015, Classical and Quantum Gravity, 32, 074001, 1411.4547

\bibitem[{{Abbott} {et~al.}(2017{\natexlab{a}}){Abbott}, {Abbott}, {Abbott},
  {Acernese}, {Ackley}, {Adams}, {Adams}, {Addesso}, {Adhikari}, {Adya}, \&
  et~al.}]{GW170817:H0}
{Abbott}, B.~P. {et~al.} 2017{\natexlab{a}}, \nat, 551, 85, 1710.05835

\bibitem[{{Abbott} {et~al.}(2017{\natexlab{b}}){Abbott}, {Abbott}, {Abbott},
  {Acernese}, {Ackley}, {Adams}, {Adams}, {Addesso}, {Adhikari}, {Adya}, \&
  et~al.}]{GW170817:GRB}
------. 2017{\natexlab{b}}, \apjl, 848, L13, 1710.05834

\bibitem[{{Abbott} {et~al.}(2017{\natexlab{c}}){Abbott}, {Abbott}, {Abbott},
  {Acernese}, {Ackley}, {Adams}, {Adams}, {Addesso}, {Adhikari}, {Adya}, \&
  et~al.}]{GW170817}
------. 2017{\natexlab{c}}, Physical Review Letters, 119, 161101, 1710.05832

\bibitem[{{Abbott} {et~al.}(2017{\natexlab{d}}){Abbott}, {Abbott}, {Abbott},
  {Acernese}, {Ackley}, {Adams}, {Adams}, {Addesso}, {Adhikari}, {Adya}, \&
  et~al.}]{GW170817:MMA}
------. 2017{\natexlab{d}}, \apjl, 848, L12, 1710.05833

\bibitem[{Acernese {et~al.}(2015)}]{AdvVirgo}
Acernese, F., {et~al.} 2015, Class. Quant. Grav., 32, 024001, 1408.3978

\bibitem[{{Alexander} {et~al.}(2017){Alexander}, {Berger}, {Fong}, {Williams},
  {Guidorzi}, {Margutti}, {Metzger}, {Annis}, {Blanchard}, {Brout}, {Brown},
  {Chen}, {Chornock}, {Cowperthwaite}, {Drout}, {Eftekhari}, {Frieman}, {Holz},
  {Nicholl}, {Rest}, {Sako}, {Soares-Santos}, \& {Villar}}]{Alexander:2017}
{Alexander}, K.~D. {et~al.} 2017, \apjl, 848, L21, 1710.05457

\bibitem[{{Bonvin} {et~al.}(2017){Bonvin}, {Courbin}, {Suyu}, {Marshall},
  {Rusu}, {Sluse}, {Tewes}, {Wong}, {Collett}, {Fassnacht}, {Treu}, {Auger},
  {Hilbert}, {Koopmans}, {Meylan}, {Rumbaugh}, {Sonnenfeld}, \&
  {Spiniello}}]{Bonvin:2017}
{Bonvin}, V. {et~al.} 2017, \mnras, 465, 4914, 1607.01790

\bibitem[{{DES Collaboration} {et~al.}(2017){DES Collaboration}, {Abbott},
  {Abdalla}, {Annis}, {Bechtol}, {Benson}, {Bernstein}, {Bernstein}, {Bertin},
  {Brooks}, {Burke}, {Carnero Rosell}, {Carrasco Kind}, {Carretero},
  {Castander}, {Chang}, {Crawford}, {Cunha}, {D'Andrea}, {da Costa}, {Davis},
  {Desai}, {Diehl}, {Dietrich}, {Doel}, {Drlica-Wagner}, {Evrard}, {Fernandez},
  {Flaugher}, {Frieman}, {Garcia-Bellido}, {Gaztanaga}, {Gerdes},
  {Giannantonio}, {Gruen}, {Gruendl}, {Gschwend}, {Gutierrez}, {Hartley},
  {Henning}, {Honscheid}, {Hoyle}, {Jain}, {James}, {Jarvis}, {Jeltema},
  {Johnson}, {Johnson}, {Krause}, {Kuehn}, {Kuhlmann}, {Kuropatkin}, {Lahav},
  {Liddle}, {Lima}, {Lin}, {Maia}, {Manzotti}, {March}, {Marshall}, {Miquel},
  {Mohr}, {Natoli}, {Nugent}, {Ogando}, {Park}, {Plazas}, {Reichardt}, {Reil},
  {Roodman}, {Ross}, {Rozo}, {Rykoff}, {Sanchez}, {Scarpine}, {Schubnell},
  {Sevilla-Noarbe}, {Smith}, {Smith}, {Soares-Santos}, {Sobreira}, {Suchyta},
  {Tarle}, {Thomas}, {Troxel}, {Walker}, {Wechsler}, {Weller}, {Wester}, {Wu},
  \& {Zuntz}}]{DES:2017}
{DES Collaboration} {et~al.} 2017, ArXiv e-prints, 1711.00403

\bibitem[{{Evans} {et~al.}(2017){Evans}, {Cenko}, {Kennea}, {Emery}, {Kuin},
  {Korobkin}, {Wollaeger}, {Fryer}, {Madsen}, {Harrison}, {Xu}, {Nakar},
  {Hotokezaka}, {Lien}, {Campana}, {Oates}, {Troja}, {Breeveld}, {Marshall},
  {Barthelmy}, {Beardmore}, {Burrows}, {Cusumano}, {D'Ai}, {D'Avanzo},
  {D'Elia}, {de Pasquale}, {Even}, {Fontes}, {Forster}, {Garcia}, {Giommi},
  {Grefenstette}, {Gronwall}, {Hartmann}, {Heida}, {Hungerford}, {Kasliwal},
  {Krimm}, {Levan}, {Malesani}, {Melandri}, {Miyasaka}, {Nousek}, {O'Brien},
  {Osborne}, {Pagani}, {Page}, {Palmer}, {Perri}, {Pike}, {Racusin}, {Rosswog},
  {Siegel}, {Sakamoto}, {Sbarufatti}, {Tagliaferri}, {Tanvir}, \&
  {Tohuvavohu}}]{Evans:2017}
{Evans}, P.~A. {et~al.} 2017, ArXiv e-prints, 1710.05437

\bibitem[{{Fong} \& {Berger}(2013)}]{FongBerger:2013}
{Fong}, W., \& {Berger}, E. 2013, \apj, 776, 18, 1307.0819

\bibitem[{{Fong} {et~al.}(2015){Fong}, {Berger}, {Margutti}, \&
  {Zauderer}}]{Fong:2015}
{Fong}, W., {Berger}, E., {Margutti}, R., \& {Zauderer}, B.~A. 2015, \apj, 815,
  102, 1509.02922

\bibitem[{{Fraija} {et~al.}(2017){Fraija}, {De Colle}, {Veres}, {Dichiara},
  {Barniol Duran}, \& {Galvan-Gamez}}]{Fraija:2017}
{Fraija}, N., {De Colle}, F., {Veres}, P., {Dichiara}, S., {Barniol Duran}, R.,
  \& {Galvan-Gamez}, A. 2017, ArXiv e-prints, 1710.08514

\bibitem[{{Granot} {et~al.}(2017){Granot}, {Gill}, {Guetta}, \& {De
  Colle}}]{Granot:2017afterglow}
{Granot}, J., {Gill}, R., {Guetta}, D., \& {De Colle}, F. 2017, ArXiv e-prints,
  1710.06421

\bibitem[{{Guidorzi} {et~al.}(2017){Guidorzi}, {Margutti}, {Brout}, {Scolnic},
  {Fong}, {Alexander}, {Cowperthwaite}, {Annis}, {Berger}, {Blanchard},
  {Chornock}, {Coppejans}, {Eftekhari}, {Frieman}, {Huterer}, {Nicholl},
  {Soares-Santos}, {Terreran}, {Villar}, \& {Williams}}]{Guidorzi:2017}
{Guidorzi}, C. {et~al.} 2017, ArXiv e-prints, 1710.06426

\bibitem[{{Haggard} {et~al.}(2017){Haggard}, {Nynka}, {Ruan}, {Kalogera},
  {Cenko}, {Evans}, \& {Kennea}}]{Haggard:2017}
{Haggard}, D., {Nynka}, M., {Ruan}, J.~J., {Kalogera}, V., {Cenko}, S.~B.,
  {Evans}, P., \& {Kennea}, J.~A. 2017, \apjl, 848, L25, 1710.05852

\bibitem[{{Hallinan} {et~al.}(2017){Hallinan}, {Corsi}, {Mooley}, {Hotokezaka},
  {Nakar}, {Kasliwal}, {Kaplan}, {Frail}, {Myers}, {Murphy}, {De}, {Dobie},
  {Allison}, {Bannister}, {Bhalerao}, {Chandra}, {Clarke}, {Giacintucci}, {Ho},
  {Horesh}, {Kassim}, {Kulkarni}, {Lenc}, {Lockman}, {Lynch}, {Nichols},
  {Nissanke}, {Palliyaguru}, {Peters}, {Piran}, {Rana}, {Sadler}, \&
  {Singer}}]{Hallinan:2017}
{Hallinan}, G. {et~al.} 2017, ArXiv e-prints, 1710.05435

\bibitem[{{Henning} {et~al.}(2017){Henning}, {Sayre}, {Reichardt}, {Ade},
  {Anderson}, {Austermann}, {Beall}, {Bender}, {Benson}, {Bleem}, {Carlstrom},
  {Chang}, {Chiang}, {Cho}, {Citron}, {Corbett Moran}, {Crawford}, {Crites},
  {de Haan}, {Dobbs}, {Everett}, {Gallicchio}, {George}, {Gilbert},
  {Halverson}, {Harrington}, {Hilton}, {Holder}, {Holzapfel}, {Hoover}, {Hou},
  {Hrubes}, {Huang}, {Hubmayr}, {Irwin}, {Keisler}, {Knox}, {Lee}, {Leitch},
  {Li}, {Lowitz}, {Manzotti}, {McMahon}, {Meyer}, {Mocanu}, {Montgomery},
  {Nadolski}, {Natoli}, {Nibarger}, {Novosad}, {Padin}, {Pryke}, {Ruhl},
  {Saliwanchik}, {Schaffer}, {Sievers}, {Smecher}, {Stark}, {Story}, {Tucker},
  {Vanderlinde}, {Veach}, {Vieira}, {Wang}, {Whitehorn}, {Wu}, \&
  {Yefremenko}}]{Henning:2017}
{Henning}, J.~W. {et~al.} 2017, ArXiv e-prints, 1707.09353

\bibitem[{{Hjorth} {et~al.}(2017){Hjorth}, {Levan}, {Tanvir}, {Lyman},
  {Wojtak}, {Schr{\o}der}, {Mandel}, {Gall}, \& {Bruun}}]{Hjorth:2017}
{Hjorth}, J. {et~al.} 2017, \apjl, 848, L31, 1710.05856

\bibitem[{{Kim} {et~al.}(2017){Kim}, {Schulze}, {Resmi},
  {Gonz{\'a}lez-L{\'o}pez}, {Higgins}, {Ishwara-Chandra}, {Bauer}, {de
  Gregorio-Monsalvo}, {De Pasquale}, {de Ugarte Postigo}, {Kann},
  {Mart{\'{\i}}n}, {Oates}, {Starling}, {Tanvir}, {Buchner}, {Campana}, {Cano},
  {Covino}, {Fruchter}, {Fynbo}, {Hartmann}, {Hjorth}, {Jakobsson}, {Levan},
  {Malesani}, {Micha{\l}owski}, {Milvang-Jensen}, {Misra}, {O'Brien},
  {S{\'a}nchez-Ram{\'{\i}}rez}, {Th{\"o}ne}, {Watson}, \&
  {Wiersema}}]{Kim:2017}
{Kim}, S. {et~al.} 2017, \apjl, 850, L21, 1710.05847

\bibitem[{{Lazzati} {et~al.}(2017){Lazzati}, {Perna}, {Morsony},
  {L{\'o}pez-C{\'a}mara}, {Cantiello}, {Ciolfi}, {giacomazzo}, \&
  {Workman}}]{Lazzati:2017}
{Lazzati}, D., {Perna}, R., {Morsony}, B.~J., {L{\'o}pez-C{\'a}mara}, D.,
  {Cantiello}, M., {Ciolfi}, R., {giacomazzo}, B., \& {Workman}, J.~C. 2017,
  ArXiv e-prints, 1712.03237

\bibitem[{{Levan} {et~al.}(2017){Levan}, {Lyman}, {Tanvir}, {Mandel},
  {et~al.}}]{Levan:2017GCN}
{Levan}, A., {Lyman}, J., {Tanvir}, N., {Mandel}, I., {et~al.} 2017, GRB
  Coordinates Network, Circular Service, No.~22207 (2017), 22207

\bibitem[{{Margutti} {et~al.}(2017){Margutti}, {Berger}, {Fong}, {Guidorzi},
  {Alexander}, {Metzger}, {Blanchard}, {Cowperthwaite}, {Chornock},
  {Eftekhari}, {Nicholl}, {Villar}, {Williams}, {Annis}, {Brown}, {Chen},
  {Doctor}, {Frieman}, {Holz}, {Sako}, \& {Soares-Santos}}]{Margutti:2017}
{Margutti}, R. {et~al.} 2017, \apjl, 848, L20, 1710.05431

\bibitem[{{Mooley} {et~al.}(2017){Mooley}, {Nakar}, {Hotokezaka}, {Hallinan},
  {Corsi}, {Frail}, {Horesh}, {Murphy}, {Lenc}, {Kaplan}, {De}, {Dobie},
  {Chandra}, {Deller}, {Gottlieb}, {Kasliwal}, {Kulkarni}, {Myers}, {Nissanke},
  {Piran}, {Lynch}, {Bhalerao}, {Bourke}, {Bannister}, \&
  {Singer}}]{Mooley:2017}
{Mooley}, K.~P. {et~al.} 2017, ArXiv e-prints, 1711.11573

\bibitem[{{Nicholl} {et~al.}(2017){Nicholl}, {Berger}, {Kasen}, {Metzger},
  {Elias}, {Brice{\~n}o}, {Alexander}, {Blanchard}, {Chornock},
  {Cowperthwaite}, {Eftekhari}, {Fong}, {Margutti}, {Villar}, {Williams},
  {Brown}, {Annis}, {Bahramian}, {Brout}, {Brown}, {Chen}, {Clemens},
  {Dennihy}, {Dunlap}, {Holz}, {Marchesini}, {Massaro}, {Moskowitz},
  {Pelisoli}, {Rest}, {Ricci}, {Sako}, {Soares-Santos}, \&
  {Strader}}]{Nicholl:2017}
{Nicholl}, M. {et~al.} 2017, \apjl, 848, L18, 1710.05456

\bibitem[{{Perego} {et~al.}(2017){Perego}, {Radice}, \&
  {Bernuzzi}}]{Perego:2017}
{Perego}, A., {Radice}, D., \& {Bernuzzi}, S. 2017, \apjl, 850, L37, 1711.03982

\bibitem[{{Pian} {et~al.}(2017){Pian}, {D'Avanzo}, {Benetti}, {Branchesi},
  {Brocato}, {Campana}, {Cappellaro}, {Covino}, {D'Elia}, {Fynbo}, {Getman},
  {Ghirlanda}, {Ghisellini}, {Grado}, {Greco}, {Hjorth}, {Kouveliotou},
  {Levan}, {Limatola}, {Malesani}, {Mazzali}, {Melandri}, {M{\o}ller},
  {Nicastro}, {Palazzi}, {Piranomonte}, {Rossi}, {Salafia}, {Selsing},
  {Stratta}, {Tanaka}, {Tanvir}, {Tomasella}, {Watson}, {Yang}, {Amati},
  {Antonelli}, {Ascenzi}, {Bernardini}, {Bo{\"e}r}, {Bufano}, {Bulgarelli},
  {Capaccioli}, {Casella}, {Castro-Tirado}, {Chassande-Mottin}, {Ciolfi},
  {Copperwheat}, {Dadina}, {De Cesare}, {di Paola}, {Fan}, {Gendre},
  {Giuffrida}, {Giunta}, {Hunt}, {Israel}, {Jin}, {Kasliwal}, {Klose}, {Lisi},
  {Longo}, {Maiorano}, {Mapelli}, {Masetti}, {Nava}, {Patricelli}, {Perley},
  {Pescalli}, {Piran}, {Possenti}, {Pulone}, {Razzano}, {Salvaterra},
  {Schipani}, {Spera}, {Stamerra}, {Stella}, {Tagliaferri}, {Testa}, {Troja},
  {Turatto}, {Vergani}, \& {Vergani}}]{Pian:2017}
{Pian}, E. {et~al.} 2017, \nat, 551, 67, 1710.05858

\bibitem[{{Planck Collaboration} {et~al.}(2016){Planck Collaboration}, {Ade},
  {Aghanim}, {Arnaud}, {Ashdown}, {Aumont}, {Baccigalupi}, {Banday},
  {Barreiro}, {Bartlett}, \& et~al.}]{Planck:2015}
{Planck Collaboration} {et~al.} 2016, \aap, 594, A13, 1502.01589

\bibitem[{{Riess} {et~al.}(2016){Riess}, {Macri}, {Hoffmann}, {Scolnic},
  {Casertano}, {Filippenko}, {Tucker}, {Reid}, {Jones}, {Silverman},
  {Chornock}, {Challis}, {Yuan}, {Brown}, \& {Foley}}]{Riess:2016}
{Riess}, A.~G. {et~al.} 2016, \apj, 826, 56, 1604.01424

\bibitem[{{Ruan} {et~al.}(2017){Ruan}, {Nynka}, {Haggard}, {Kalogera}, \&
  {Evans}}]{Ruan:2017}
{Ruan}, J.~J., {Nynka}, M., {Haggard}, D., {Kalogera}, V., \& {Evans}, P. 2017,
  ArXiv e-prints, 1712.02809

\bibitem[{{Troja} {et~al.}(2017){Troja}, {Piro}, {van Eerten}, {Wollaeger},
  {Im}, {Fox}, {Butler}, {Cenko}, {Sakamoto}, {Fryer}, {Ricci}, {Lien}, {Ryan},
  {Korobkin}, {Lee}, {Burgess}, {Lee}, {Watson}, {Choi}, {Covino}, {D'Avanzo},
  {Fontes}, {Gonz{\'a}lez}, {Khandrika}, {Kim}, {Kim}, {Lee}, {Lee}, {Kutyrev},
  {Lim}, {S{\'a}nchez-Ram{\'{\i}}rez}, {Veilleux}, {Wieringa}, \&
  {Yoon}}]{Troja:2017}
{Troja}, E. {et~al.} 2017, \nat, 551, 71, 1710.05433

\bibitem[{{Veitch} {et~al.}(2012){Veitch}, {Mandel}, {Aylott}, {Farr},
  {Raymond}, {Rodriguez}, {van der Sluys}, {Kalogera}, \&
  {Vecchio}}]{Veitch:2012}
{Veitch}, J. {et~al.} 2012, \prd, 85, 104045, 1201.1195

\end{thebibliography}

\end{document}